# CBHRP: A CLUSTER BASED ROUTING PROTOCOL FOR WIRELESS SENSOR NETWORK

M. G. Rashed[1], M. Hasnat Kabir[2], M. Sajjadur Rahim[3], and Sk. Enayet Ullah[2]

[1]Department of Electronics Telecommunication Engineering,
Prime University, Dhaka, Bangladesh
golamrashed@primeuniversity.edu.bd
[2]Department of Information and Communication Engineering,
University of Rajshahi, Rajshahi, Bangladesh
{hasnatkabir,enayet67}@yhaoo.com
[3]School of Information Engineering, University of Padova, Italy,
sajid83@yhaoo.com

***ABSTRACT***

*A new two layer hierarchical routing protocol called Cluster Based Hierarchical Routing Protocol (CBHRP) is proposed in this paper. It is an extension of LEACH routing protocol. We introduce cluster head-set idea for cluster-based routing where several clusters are formed with the deployed sensors to collect information from target field. On rotation basis, a head-set member receives data from the neighbor nodes and transmits the aggregated results to the distance base station. This protocol reduces energy consumption quite significantly and prolongs the life time of sensor network. It is found that CBHRP performs better than other well accepted hierarchical routing protocols like LEACH in term of energy consumption and time requirement.*

## 1. INTRODUCTION

Wireless sensor networks (WSN) have been identified as one of the most important technologies in the 21st century for various applications such as habitat monitoring, automation, agriculture, and security. WSN consists of several tiny sensors called nodes that are organized in spatially distributed topography. Researchers define WNS as an important component in the field of Ubiquitous computing [1, 2]. Proactive computing concepts apply on WSN. With the proactive computing model, computers will anticipate our needs and sometimes take action on our behalf. Wireless sensor networks and proactive computing can assist us to improve efficiency, have data from places which are otherwise difficult to get or to costly to monitor [3].

The dimension of microsensor is another important design goal. The size of sensor becomes very tiny day by day. The nodes are typically battery operated sensing devices with limited energy resources. Due to the small size, the battery of the nodes is fixed. One of the elegant ways to turn off the sensor hardware when they are not used [4]. But replacing the sensor nodes is difficult when their energy is depleted [5].

Scalability and performance consistency are the major design attributes of sensor networks. To allow the system to cope with additional load and to be able to cover a large area of interest without degrading the services, networking clustering has been pursued in some routing approach [6]. The main objectives of clustering hierarchy is to efficiently maintain the energy consumption of sensor nodes by involving them in ad-hoc communication within a particular cluster and by performing data aggregation and fusion in order to decrease the number of transmitted message to the base station. Cluster formation is typically based on the reserve of sensors and sensors' proximity to the cluster head.

The lifetime of sensor can be increased by optimized the applications, operating systems, and communication protocols. Using existing hardware, improvement of communication protocol can prolong the lifetime of the network. This feature of WSN has opened new era for scientists and engineers to develop efficient routing protocol. Numerous concept of communication protocol can be found in literature [7, 8] to reduce the energy consumption of the wireless sensor networks. Low-Energy Adaptive Clustering Hierarchy (LEACH), is one of the best communication protocols for wireless sensor networks [9-11]. However, improvement of LEACH protocol is going on to make it more efficient [4,], [12-13]. Therefore, it is significant to extend the sensor networks lifetime through proficient use of the sensor nodes energy. Our new proposed protocol increases the lifetime effectively than that of others by reducing the energy consumption.

## 2. RELATED WORKS

The key responsibility of wireless sensor network is to forward the sensing data gathered by sensor nodes in the sensing fields to the BS. One simple approach to the fulfilment of this task is direct data transmission. In this case, each node in the network directly sends sensing data to the BS. As observed in LEACH [9], the direct approach would work best if the BS is located close to the sensor nodes or the cost of receiving is very high as compared to the cost of transmitting data. However, if the BS is remote from the sensor node, the node will soon die for suffering excessive energy consumption for delivering data. To solve this problem, some algorithms that are aimed to save energy have been proposed one after another.

LEACH protocol is hierarchical routing algorithm that can organize nodes into clusters collections. Each cluster controlled by cluster head. Cluster head has several duties. First one is gathering data from member cluster and accumulates them. Second one is directly sending accumulation data to sink. Third one is scheduling based of Time-Division Multiple Access (TDMA). In that, each node in cluster related to its time slot could send collection data [9]. Cluster head announce time slot by uses of distribution property to all members of cluster. Main operations of LEACH are classify in two separate phase. First phase or initialization phase has two process; clustering and cluster head determining. Second phase mean steady-state, that this phase concentrate to gathering, accumulation and transmit data to sink. BCDCP [10] is a dynamic clustering protocol, which distributes the energy dissipation evenly among all sensor nodes to improve network lifetime and average energy savings. Simulation results show that BCDCP reduces overall energy consumption and improves network lifetime over LEACH, LEACH-C [11], and PEGASIS. HEARP [16] is a hierarchical energy-efficient routing protocol. HEARP is based on both LEACH and PEGASIS [17] protocols. In HEARP, network establishment begins with the formation of clusters. Several clusters are formed with one cluster head (CH) in each cluster. Each cluster contains several nodes called member nodes, after the clusters are formed; a chain is established among all the CHs using a greedy algorithm. A CH is chosen as leader node form this chain for sending data to the BS. The operation of HEARP is broken up into rounds, where each round begins with a set-up phase, followed by data transmission phase. During the set-up phase, at the beginning of cluster formation, HEARP makes use of the same algorithm as LAECH. It has been found from simulation results that HEARP is better than LEACH, in terms of energy consumption. Again in terms of latency, HEARP performs better than LEACH as well as PEGASIS. HEARP saves energy because only one node transmits data directly to the base station.

## 3. THE PROPOSED APPROACH (CBHRP)

We have proposed and designed Cluster Based Hierarchical Routing Protocol (CBHRP) as an extension of LEACH. It is a two-layer protocol where a number of clusters cover the whole region of the network. The presented protocol is an optimum energy efficient cluster based hierarchical routing protocol for wireless sensor network. Here we introduce a concept of head-set instead of only head for cluster-based routing. The head-set members are responsible for the control and the management of network. At a time, only one member of the head-set is active and the remaining are in sleep mode within a cluster. This protocol divides the network into a few real clusters that are managed by a virtual cluster head. The tasks are uniformly distributed among all the head-set members. For a given number of data collecting nodes, the head-set nodes are systematically adjusted to reduce the energy consumption, which increases the network life. A detail architectural description of this proposed protocol can be found in our previous work [18].

## 4. SIMULATION MODEL OF CBHRP

We use a similar radio model as described in [9] where for a short distance transmission, such as within a clusters, the energy consumed by a transmit amplifier is proportional to d2. However, for a long distance transmission, such as from a cluster head to the base station, the energy consumption is proportional to d4 .Using given radio model, the energy consumption to transmit an l-bits message for a long distance (CH to BS), is given by:

$$E_T = lE_{elec} + l\varepsilon_{amp-l} d^4, \quad (1)$$

Similarly, the energy consumption to transmit an l-bits message for a short distance is given by:

$$E_T = lE_{elec} + l\varepsilon_{amp-s} d^2, \quad (2)$$

However, the energy consumption to receive the l-bits message is given by:

$$E_R = lE_{elec} + lE_{AG}, \quad (3)$$

Eq. (3) includes the data aggregation approach.

### 4.1. Election Phase for CBHRP

Consider all nodes have same energy. Therefore, the amount of consumed energy is same for the entire cluster. For uniformly distributed clusters, each cluster contains $\frac{n}{k}$ nodes. Using Eq. (2) and Eq. (3), the energy consumed by a cluster head is estimated as follows:

$$E_{CH-election} = \{lE_{elec} + l\varepsilon_{amp-s} d^2\} + \{(\tfrac{n}{k} - 1) lE_{elec} + lE_{AG}\} \quad (4)$$

The first part of Eq. (4) represents consumed energy to transmit the advertisement message; this energy consumption is based on short distance energy dissipation model. The second part represents consumed energy to receive $(\frac{n}{k} - 1)$ messages from the normal sensor nodes of the same cluster. Eq. (4) can be simplified as follows:

$$E_{CH-election} = (\tfrac{n}{k})lE_{elec} + lE_{AG}(\tfrac{n}{k}-1) + l\varepsilon_{amp-s}d^2 \quad (5)$$

Using Eq. (2) and Eq. (3), the energy consumed by non-cluster head sensor nodes are estimated as follows:

$$E_{non-CH-election} = \{klE_{Elec} + klE_{AG}\} + \{lE_{elec} + l\varepsilon_{amp-s}d^2\} \quad (6)$$

The first part of Eq. (6) shows consumed energy to receive messages from k cluster heads; it is assumed that a sensor node receives messages from all the cluster heads. The second part shows consumed energy to transmit the decision to the corresponding cluster head. Eq. (6) can be simplified as follows:

$$E_{non-CH-election} = lE_{elec}(1+k) + klE_{AG} + l\varepsilon_{amp-s}d^2 \quad (7)$$

## 4.2. Data transfer phase for CBHRP

During data transfer phase, the nodes transmit messages to their cluster head and cluster heads transmit an aggregated message to a distant base station. The energy consumed by a cluster head is as follows:

$$E_{CH/frame} = \{lE_{elec} + l\varepsilon_{amp-l}d^4\} + \{(\tfrac{n}{k}-m)l(E_{elec} + E_{AG})\} \quad (8)$$

The first part of Eq. (8) shows consumed energy to transmit a message to the distant base station. The second part shows consumed energy to receive messages from the remaining $(\tfrac{n}{k}-1)$ nodes. Eq. (8) can be simplified as follows:

$$E_{CH/frame} = l\varepsilon_{amp-l}d^2 + (\tfrac{n}{k}-m+1)lE_{elec} + l(\tfrac{n}{k}-m)E_{AG} \quad (9)$$

The energy, $E_{non-CH/frame,}$ consumed by a non-cluster head node to transmit the sensor data to the cluster head is given below:

$$E_{non-CH/frame} = lE_{elec} + l\varepsilon_{amp-s}d^2 \quad (10)$$

In one iteration, $N_f$ data frames are transmitted. Each cluster transmits $N_f/k$ data frame. These data are uniformly divided among n/k nodes. Therefore, a cluster head transmits the data frame of (n/k-m) non-cluster head data. For simplification of equations, the fraction $f_1$ and $f_2$ are given as below:

$$f_1 = \left\{\frac{1}{\tfrac{n}{k}-m+1}\right\} * \frac{1}{k} \quad (11)$$

$$f_2 = \left\{ \frac{\frac{n}{k} - m}{\frac{n}{k} - m + 1} \right\} * \frac{1}{k} \tag{12}$$

The energy consumption in a data transfer phase of each cluster is as follows:

$$E_{CH-data} = f_1 N_f E_{CH/frame} \tag{13}$$

$$E_{non-CH-data} = f_2 N_f E_{non-CH/frame} \tag{14}$$

### 4.3. Initial Energy for one round in CBHRP

In each iteration, m nodes are elected as associate head for each cluster. Thus, km nodes are elected as members of head-sets. The numbers of iterations required in one round are $\frac{n}{km}$ that are needed to be elected for all nodes. Each iteration consists of an election phase and a data transfer phase. Therefore, the consumed energy in an iteration is as follows:

$$E_{CH/iteration/cluster} = E_{CH-election} + E_{CH-data} \tag{15}$$

$$E_{non-CH/iteration/cluster} = E_{non-CH-election} + E_{non-CH-data} \tag{16}$$

Since, there is m nodes as cluster head within a cluster, the ECH/iteration/cluster is assigned for all cluster heads. For single cluster head is

$$E_{CH/node} = \frac{E_{CH/iteration/cluster}}{m} \tag{17}$$

Similarly, there are ( $\frac{n}{k} - m$ ) non-cluster head nodes in a cluster. The Enon-CH/iteration/cluster is uniformly distributed among all the non-cluster head members. For single non-cluster head is

$$E_{non-CH/node} = \frac{E_{non-CH/iteration/cluster}}{(\frac{n}{k} - m)} \tag{18}$$

The initial energy, $E_{init}$ should be sufficient for at least one round. A node becomes a member of head-set for one time and a non-cluster head for (n/mk-1) times in one round. An estimation of $E_{init}$ is given below:

$$E_{init} = E_{CH/node} + (\frac{n}{mk} - 1) E_{non-CH/node} \tag{19}$$

Eq. (19) can be represented as follows using the value of Eq. (17) and Eq. (18)

$$E_{init} = \frac{E_{CH/iteration/cluster} + E_{non-CH/iteration/cluster}}{m} \tag{20}$$

Using Eq. (13), (14), (15), (16) Einit can be given as follows:

$$E_{init} = \left\{\frac{E_{CH-election} + E_{non-CH-election}}{m}\right\} + \left\{\frac{N_f}{m} * \left(f_1 E_{CH/frame} + f_2 E_{non-CH/frame}\right)\right\} \tag{21}$$

Using Eq. (21), Nf can be given as follows:

$$N_f = \frac{mE_{init} - E_{CH-election} - E_{non-CH-election}}{f_1 E_{CH/frame} + f_2 E_{non-CH/frame}} \tag{22}$$

## 4.4. Required Time for one round

Sensor nodes transmit messages according to a specified schedule, which is based on TDMA. The frame time, $t_{frame}$ is the integration of frame transmitted time by all the nodes of a cluster. For a data transfer rate of $R_b$ bits/second and message length of l bits, the time to transfer a message, $t_{msg}$, is:

$$t_{msg} = \frac{l}{R_b} \tag{23}$$

The messages are transmitted by all the non-cluster head nodes and the active member of cluster head-set. Since at one time only one member of head-set is active, the remaining (m-1) inactive head-sets do not transmit any frame.

The time for one frame is

$$t_{frame} = \sum_{i=1}^{\frac{n}{k}-m} t_{msgi} + t_{msgcluster_head} \tag{24}$$

The first part of Eq. (24) is due to ( $\frac{n}{k} - m$ ) message from non-cluster head nodes. The second part is due to the active member of the head-set.

If we assume that message transfer time is same for all the nodes, Eq. (24) can be simplified as follows:

$$t_{frame} = (\frac{n}{k} - m + 1)t_{msg} \tag{25}$$

As $N_f$ frames are transmitted in one epoch, time for one iteration, $t_{iteration}$ is:

$$t_{iteraton} = t_{frame} \times N_f \tag{26}$$

Using Eq. (23), (25), (26), the iteration time $t_{iteration}$ can be given as:

$$t_{iteration} = \frac{l}{R_b}(\frac{n}{k} - m + 1)N_f \tag{27}$$

Since there are $\frac{n}{km}$ iterations in one round, the time for one round, tround is:

$$t_{round} = t_{iteration} \times \frac{n}{km}$$

$$= \frac{l}{R_b}\frac{n}{k}(\frac{n}{k} - m + 1)\frac{N_f}{m} \tag{28}$$

## 4. RESULTS AND DISCUSSION

In wireless sensor network, energy is prime concerned factor. Low energy consume network is desirable. The variation of energy consumption per round with respect to the number of clusters and network diameter is shows in Figure 1. Graph represents that energy consumption is reduced when the number of clusters are increased. It also shows that the optimum range of cluster is from 20 to 60. When the number of clusters are below the optimum range, for example 10, normal sensor nodes have to send data to the distance cluster heads with high-energy dissipation. On the other hand, when the number of cluster is excited optimum range, more transmissions is possible to the distance base station with minimum energy dissipation.

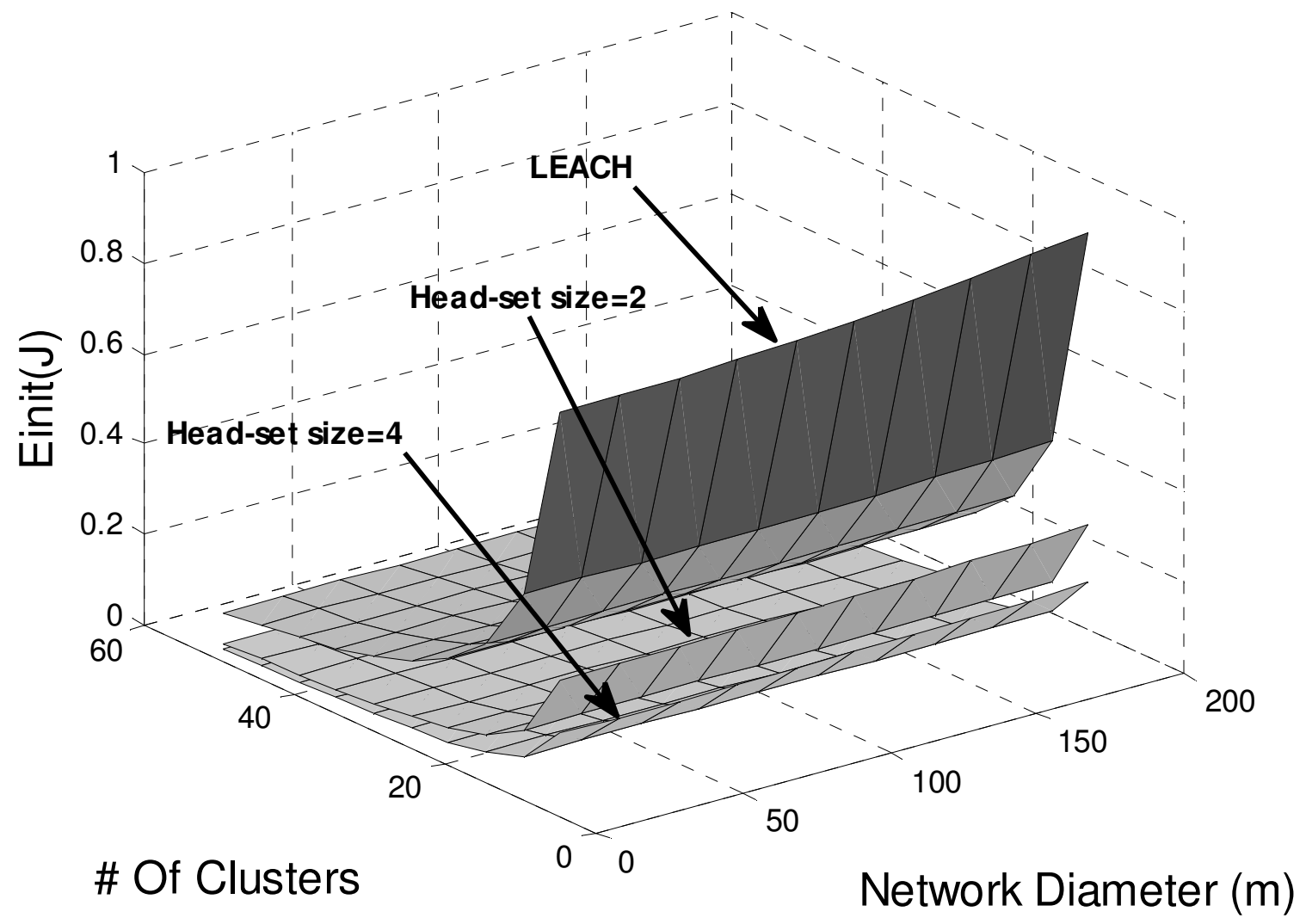

Figure 1. Energy consumption per round with respect to number of cluster and network diameter.

It can be pointed out that the concept of head-set considered in the network reduces energy consumption. In comparison with LEACH, it is clearly seen for transmission of each frame that the energy consumption is reduced by 5 and 7 times for head-set size of 2 and 4, respectively. The head-set reduces the election process to become a cluster head in our proposed routing model as compared to LEACH. The number of elections for n nodes in the case of LEACH is $\frac{n}{k}$ whereas $\frac{n}{mk}$ for CBHRP here m is the head-set size.

Figure 2. illustrates the energy consumption per round with respect to the head-set size and distance of the base station for fixed network diameter. Here the number of clusters k=50.It is seen from this figure that the energy consumption is increased with the distance of the base station for a given value of head-set. The position of the base station has great influence on the lifetime of the network. Since most of the energy is consumed by transmission from cluster-heads to the base station, therefore, energy savings are not significant for a longer distance between the base station and the sensor nodes.

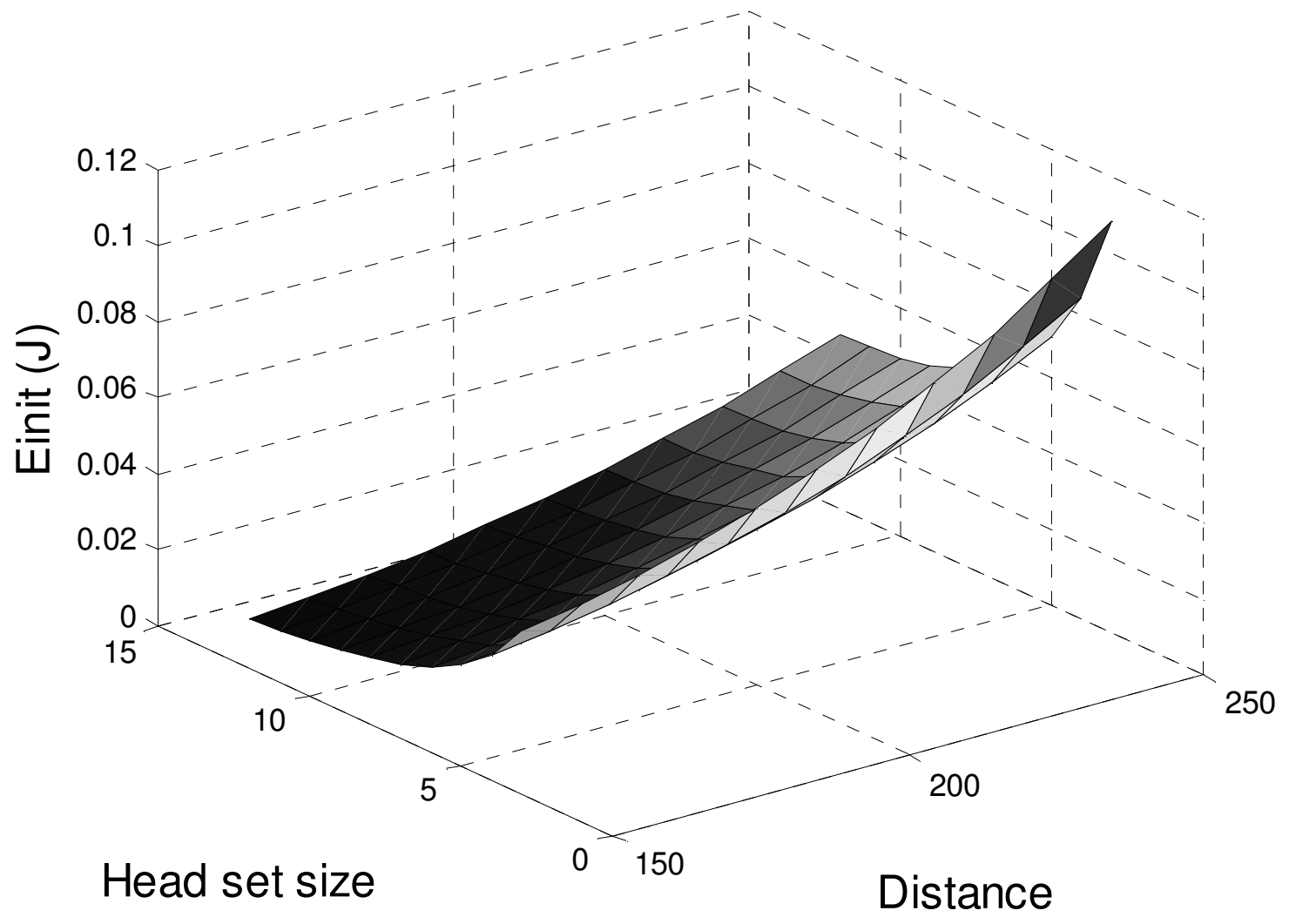

Figure 2. Energy consumption per round with respect to the distance of the BS and head set size when the network diameter is fixed.

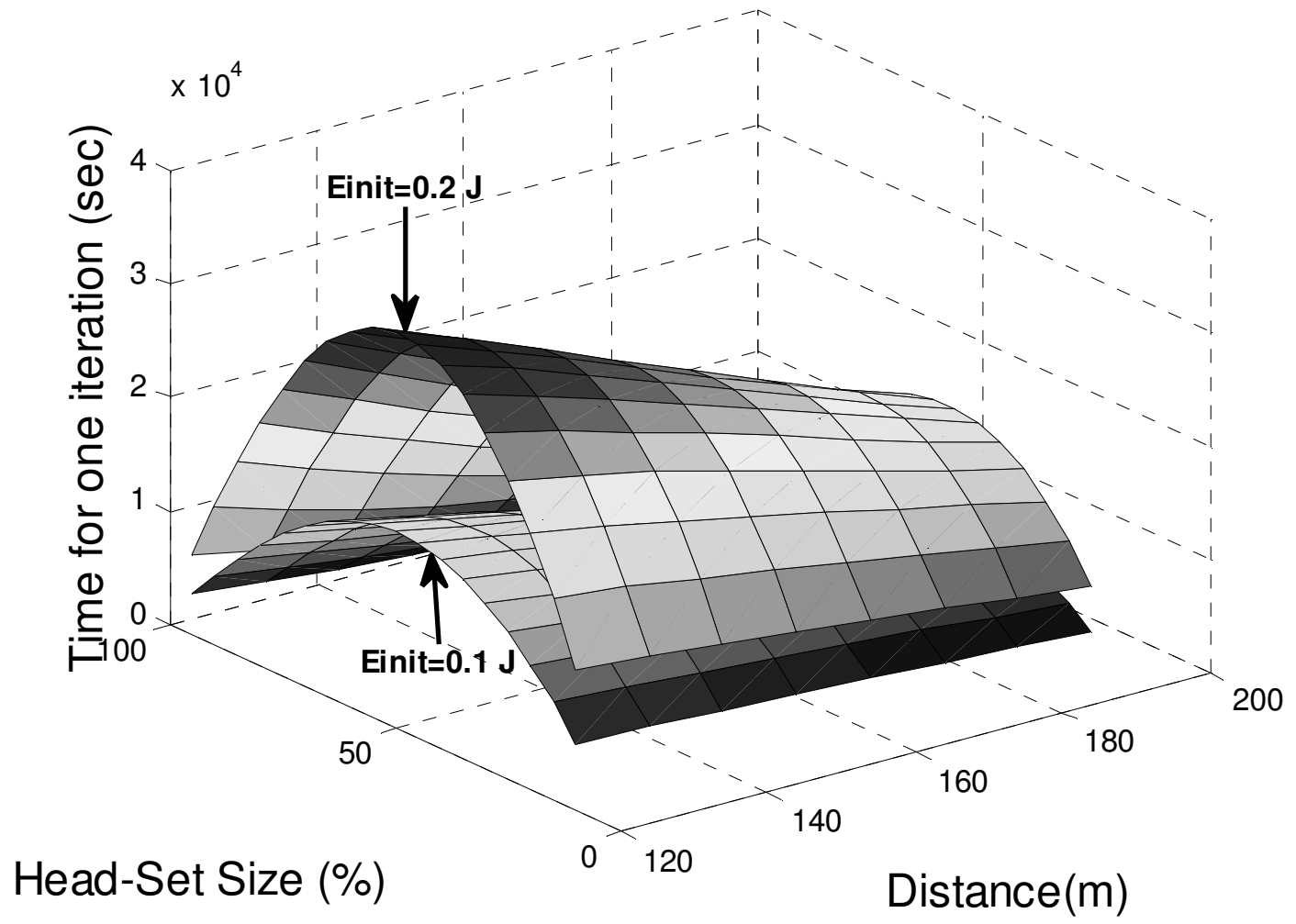

Figure 3. Iteration time with respect to distance of the BS and the head-set size.

The relationship between iteration time and the head-set size as a function of distance of base station is shown in Figure 3. For a given number of head-set, the distance of the base station may be varied to observe the impact in iteration time. The head-set size is given as a percentage of cluster size. The initial energy can be used for the longest period of time when the head set size is 50% of the cluster size. Here we see that the iteration time is decreasing with increasing of head-set size as a function of base station distance.

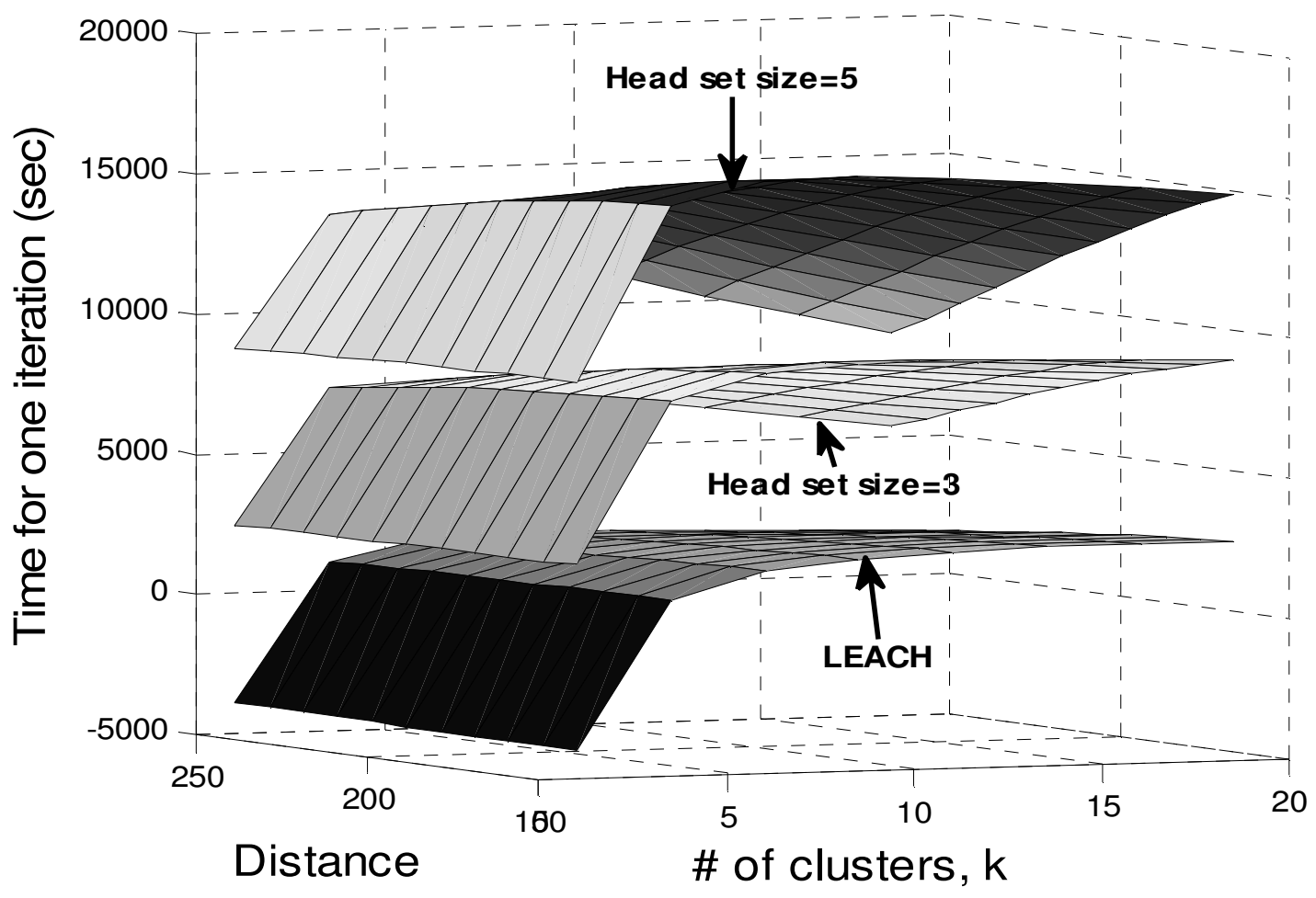

Figure 4. Iteration time with respect to the number of clusters and the distance of the BS.

Figure 4 shows a graph that illustrates the variation in time for iteration with respect to the number of clusters and BS distance. The time for iteration increases as the head-set size increases for a particular number of clusters. This indicates that the head-set size should be carefully chosen to extend the network life time.

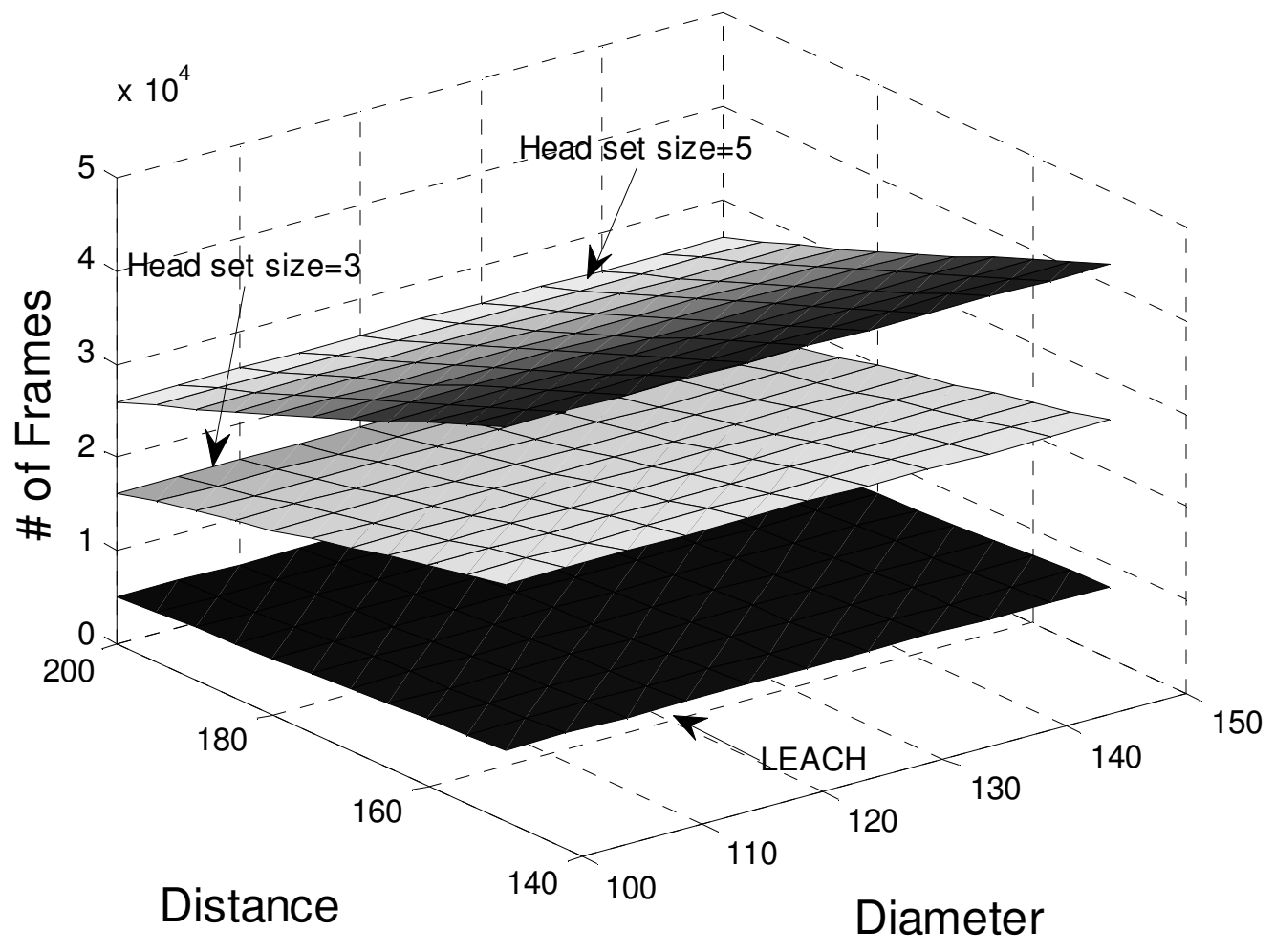

Figure 5. Number of frames transmission per iteration with respect to the distance of BS and network diameter.

Figure 5 shows the number of frames transmitted with respect to network diameter and distance from BS. The number of data transmission is another important issue in a wireless sensor network. It is a measuring tool of a network suitability and perfectness. Simulated results show that the proposed method is able to transmit higher data frame in contrast to LECAH for the same dimension and distance of BS. However, it is clearly found that the transmission rate slightly decreases with respect to the distance of BS.

## 5. CONCLUSIONS

The proposed routing protocol (CBHRP) is designed to overcome some presents limitations of WSNs. The results of our analysis indicate that the energy consumption can be considerably reduced by including more sensors as cluster head in a head-set instead of using only one cluster head within a cluster. The iteration time and data transfer rate also increases with the size of head-set. Therefore, the overall network lifetime is prolonged. So, it can be concluded that the propose protocol provides an energy-efficient routing scheme is suitable for a vast range of sensing applications than that of LECAH.